\documentstyle[11pt,epsfig]{article}
\newcommand{\ba}{\begin{array}}
\newcommand{\ea}{\end{array}}
\newcommand{\bd}{\begin{displaymath}}
\newcommand{\ed}{\end{displaymath}}
\newcommand{\be}{\begin{equation}}
\newcommand{\ee}{\end{equation}}
\newcommand{\bea}{\begin{eqnarray}}
\newcommand{\eea}{\end{eqnarray}}
\newcommand{\Dir}{\kern -6.4pt\Big{/}}
\newcommand{\Dirin}{\kern -10.4pt\Big{/}\kern 4.4pt}
\newcommand{\DDir}{\kern -10.6pt\Big{/}}
\newcommand{\DGir}{\kern -6.0pt\Big{/}}
\begin{document}
\
\def\bra{\langle}
\def\ket{\rangle}

\def\a{\alpha}
\def\as {\alpha_s}
\def\b{\beta}
\def\d{\delta}
\def\e{\epsilon}
\def\ve{\varepsilon}
\def\l{\lambda}
\def\m{\mu}
\def\n{\nu}
\def\G{\Gamma}
\def\D{\Delta}
\def\L{\Lambda}
\def\s{\sigma}
\def\p{\pi}

\def\etal{ {\em et al.}}
\def\mzs {M_Z^2}
\def\mws {M_W^2}
\def\q2 {q^2}
\def\sz {\sin^2\theta_W}
\def\cz {\cos^2\theta_W}
\def\lp{\lambda^{\prime}}
\def\lps{\lambda^{\prime *}}
\def\lpp{\lambda^{\prime\prime}}
\def\lpps{\lambda^{\prime\prime * }}

\def\bapp{b_1^{\prime\prime}}
\def\bbpp{b_2^{\prime\prime}}
\def\bcp{b_3^{\prime}}
\def\bdp{b_4^{\prime}}
\def\t {\times }
\def\slash {\!\!\!\!\!\!/}
\def\photino {\tilde\gamma}
\def\sel {\tilde{e}}
 \def\N10{\widetilde \chi_1^0}
                         \def\C1p{\widetilde \chi_1^+}
                         \def\C1m{\widetilde \chi_1^-}
                         \def\C1pm{\widetilde \chi_1^\pm}
 \def\Ntwo{\widetilde \chi_2^0}
                         \def\Ctwo{\widetilde \chi_2^\pm}
\def\lslep {{\tilde e}_L}
\def\rslep {{\tilde e}_R}
\def\sneu {\tilde \nu}
\def\msneu {M_\tilde \nu}
\def\mrslep {m_{\rslep}}
\def\mlslep {m_{\lslep}}
\def\mneu {m_{\neu}}
\def\mpT{p_T \hspace{-1em}/\;\:}
\def\mET{E_T \hspace{-1.1em}/\;\:}
\def\mE{E \hspace{-.7em}/\;\:}
\def\go{\rightarrow}
\def\beq{\begin{eqnarray}}
\def\Rp{R\!\!\!\!/}
\def\wrp {{\cal W}_{R\!\!\!\!/}}
\def\enq{\end{eqnarray}}
\def\goes{\longrightarrow}
\def\lsim{\:\raisebox{-0.5ex}{$\stackrel{\textstyle<}{\sim}$}\:}
\def\gsim{\:\raisebox{-0.5ex}{$\stackrel{\textstyle>}{\sim}$}\:}
 \begin{flushright}
{\large SHEP-03-16}\\
{\large July 2003}\\
{\large Revised January 2004}\\
\end{flushright}
\begin{flushleft}
\end{flushleft}
\begin{center}
{\Large\bf Detection of heavy charged Higgs bosons}\\[5mm]
{\Large\bf in $e^+e^-\to t\bar b H^-$ production}\\[5mm]
{\Large\bf at future Linear Colliders}\\[15mm]
{\Large S. Moretti\footnote{stefano@hep.phys.soton.ac.uk}}\\[4mm]
{\em School of Physics \& Astronomy, University of Southampton,\\
Highfield, Southampton SO17 1BJ, UK}
\\[20mm]
\end{center}
\begin{abstract}
\noindent
Heavy charged Higgs bosons ($H^\pm$) of a Type II 2-Higgs Doublet Model
(2HDM)  can be detected at future electron-positron Linear Colliders 
(LCs) even when their mass is larger than half the collider energy.
The single Higgs mode $e^+e^-\to t\bar b H^- + ~{\rm{c.c.}}
\to 4b +{\rm{j}}{\rm{j}}
+ \ell + p_T^{\rm{miss}}$ (where j represents a jet and
with $\ell=e,\mu$) contributes 
to extend the discovery reach of $H^\pm$ states into the mass region
$M_{H^\pm}\gsim \sqrt s/2$, where the well studied pair production channel 
$e^+e^-\to H^-H^+$ is no
longer available. With a technique that allows one to reconstruct the
neutrino four-momentum in the decay $t\to b W^+\to b \ell^+\nu$,
one can suppress the initially overwhelming main 
irreducible background due to $e^+e^-\to t\bar t b\bar b$ (via
a gluon splitting into $b\bar b$ pairs)
to a negligible level. However, for currently foreseen luminosities, 
one can establish a statistically significant $H^\pm$  signal only over
a rather limited mass region, of 20 GeV or so, beyond 
$M_{H^\pm}\approx \sqrt s/2$, 
for very large or very small values of $\tan\beta$
and provided high $b$-tagging efficiency can be achieved. 
\end{abstract}

\vskip 1 true cm

\noindent
Keywords: {Beyond Standard Model, Two Higgs Doublet Models, Charged Higgs 
Bosons}

\newpage
\noindent
Charged Higgs bosons
 appear in the particle spectrum of a general 2HDM. 
Embedding a Type II such model into the theoretical 
framework provided by Supersymmetry (SUSY) yields 
the Minimal Supersymmetric Standard Model (MSSM). In presence
of SUSY, the mass of the two charged Higgs states of the theory is
closely tied to that of the CP-odd neutral Higgs boson, denoted by $A$, 
and to those of the two CP-even neutral states, labelled as
$h$ and $H$ (in increasing order of mass). These five
states make up the Higgs particle
spectrum of a 2HDM. The ratio of the vacuum expectation values
(VEVs) of the two Higgs doublets, hereafter $\tan\beta$,  together with
the mass of one of the physical Higgs states (say, $M_A$) uniquely
defines the production and decay phenomenology of the MSSM Higgs sector at 
tree-level, provided the mass of the SUSY partners of ordinary matter
(so-called sparticles) is significantly higher than the hard scale
involved in the Higgs processes considered.

There exists a significant region of the MSSM parameter space, the so-called
`decoupling limit', namely, when $M_A\sim M_H\sim M_{H^\pm}\gg M_h$, for 
values of $\tan\beta$ between, say, 2--3 and 30--40 (the larger $M_A$ the
higher the upper limit in $\tan\beta$), where only the 
light (below 130 GeV or so) neutral Higgs boson $h$ is
found at the Large Hadron Collider (LHC) and this is degenerate 
with the SM Higgs state.
Under these circumstances, it would be very difficult to investigate
at the LHC the mechanism of Electro-Weak Symmetry Breaking (EWSB) and 
understand whether the latter is generated within the SM or else by the 
MSSM dynamics.

The availability of $e^+e^-$ LCs operating at the TeV scale or above
 \cite{eebiblio} will then be crucial to solve this puzzle. In this
respect, the accepted wisdom is that high precision measurements  can
easily be performed in such a clean environment, enabling one to asses the true
nature of such a light Higgs state, possibly inferring also the values of  
$M_A, M_H$ and $M_{H^\pm}$. In fact, mass relations among
the five MSSM Higgs states are now known very accurately, 
at the two-loop level \cite{2loop}\footnote{Recall that some virtual 
sparticle effects can enter such relations, even for high SUSY mass 
values. However, the dependence is rather mild (logarithmic, to be
precise) and almost invisible for the case of the charged Higgs
state, for which one may safely adopt the tree-level expression
$M_{H^\pm}^2=M_A^2+M_{W^\pm}^2$.}, 
as a function of $\tan\beta$, that could also be determined
rather easily at LCs.

However, it may well turn out that the extrapolated mass for the heavy
Higgs states of the model is larger than half the LC energy: i.e.,
$M_A\sim M_H\sim M_{H^\pm}\gsim \sqrt s/2$. This would be a rather 
difficult configuration to investigate even at LCs. In fact, this occurs when
the couplings $Z A h$ and $ZZH$ are minimal, hence preventing one 
from exploiting the $e^+e^-\to Ah$ 
and                 $e^+e^-\to ZH$
production processes\footnote{There is no $ZZA$ coupling at tree level in the
MSSM, hence the $e^+e^-\to ZA$ channel is phenomenologically irrelevant.} 
to access the CP-odd and heavy CP-even neutral Higgs bosons.
 The only means of producing these objects
would be via $e^+e^-\to AH$, whose cross section is maximal in the
decoupling limit, yet negligible if $M_H+ M_A\gsim \sqrt s/2$.
Similarly, the leading production mode of charged Higgs states is 
via $H^\pm$ pairs, $e^+e^-\to H^-H^+$ \cite{HpHm}, 
which presents the same drawbacks.

Clearly, the discovery of the additional Higgs states 
expected in the model (other than $h$) and the measurement of their
quantum numbers is a necessary ingredient
to definitely pin down the dynamics of the underlying EWSB
mechanism. Despite the difficult situation outlined above, 
there are two strategies that one can pursue at future
LCs. One can either exploit the $\gamma\gamma$ option to produce heavy
neutral Higgs states in single mode, via (triangle) loops of heavy fermions
($\tau$-leptons, but  chiefly $b$- and $t$-quarks), plus possibly 
$W^\pm,H^\pm$ bosons (limitedly to the CP-even state), or resort to
$e^+e^-$ production modes of $A,H$ and $H^\pm$ states that only involve one
such particles at a time. While the option of allowing for photon-photon
interactions may certainly be viable (at an energy and luminosity close
to those of the primary $e^+e^-$ design), the case for the investigation
of the second alternative is certainly stronger. (Besides, the $\gamma\gamma$
mode would not be helpful in the case of charged Higgs bosons.)

Heavy neutral Higgs bosons could for example 
be produced via $e^+e^-\to b\bar b A$
and $e^+e^-\to b\bar b H$ \cite{bbHiggs}, provided 
$\tan\beta$ is significantly above unity. Alternatively, one could try 
exploiting loop-production of $A$ bosons via 
$e^+e^-\to \nu\bar\nu A$ \cite{nunuA}. Some investigations
of these channels exist in literature, yet a dedicated
signal-to-background analysis is still missing to date.
We will address this in a separate publication \cite{preparation}.
Here, we will concentrate on single production of charged Higgs 
bosons\footnote{For similar studies in the case of $\gamma\gamma$,
or even $e\gamma$, collisions, see Refs.~\cite{2gamma,egamma}.}.

There exist several channels yielding only one $H^\pm$ state in the final state
of electron-positron annihilations at TeV energy scales,
see Refs.~\cite{singleHpm}--\cite{WH}.
The only ones though
 that can offer some chances of detection are the following:
\begin{eqnarray}
e^+e^- &\to& \tau^-\bar\nu_\tau H^+, \tau^+\nu_\tau H^-~~
\mathrm{(tree\ level)},
\label{proc_tau} \\
e^+e^- &\to& b\bar t H^+, t\bar b H^-~~
\mathrm{(tree\ level)},
\label{proc_t} \\
e^+e^- &\to& W^\mp H^\pm ~~\mathrm{(one\ loop)}.    
\label{proc_wh}
\end{eqnarray}
The first one is relevant only in the large $\tan\beta$ region, whereas the
latter is important only for the low one. The second one can cover both,
yet is of little use for intermediate $\tan\beta$ values (say, around 6--10).
As LEP2 data seem to prefer large values 
of $\tan\beta$, at least in the MSSM \cite{LEPTRE}, some 
attempt of disentangling the first process from the background
in the mass interval $M_{H^\pm}\gsim \sqrt s/2$ were carried out in
Ref.~\cite{taunu}, not without success.
In fact, some coverage was claimed over a region extending for about 
20--30 GeV above the kinematic limit  $M_{H^\pm}=\sqrt s/2$, 
for $\tan\beta\gsim30$--40. We attempt here to
devise a selection procedure that may help to extract process 
(\ref{proc_t}) from the irreducible background\footnote{We defer
a similar study of channel (\ref{proc_wh}) to a forthcoming
publication \cite{newpreparation}.}. Since, as shown in
\cite{singleHpm}, the production rates for channel (\ref{proc_t})
are rather small in general over the mass region $M_{H^\pm}\gsim\sqrt s/2$ 
(for sake of illustration, we adopt here $\sqrt s\equiv
E_{\rm{cm}}=1000$ GeV), it
is mandatory to resort to the main decay channel of heavy charged Higgs
bosons, i.e., $H^+\to t\bar b$. Hence, the following processes are of
relevance for the signal ($S$) and the main irreducible background
($B$)\footnote{Charged 
conjugated (c.c.) channels are assumed throughout the paper.}:
\begin{eqnarray}\label{signal}
e^+ e^- &\rightarrow& b \bar t H^+ + t\bar b H^-\to 
t \bar t b\bar b~~~({\mathrm{signal}}),\\ 
\label{background}
e^+ e^- &\rightarrow& t\bar t  g^*\to t\bar t b\bar b  
~~~~~({\mathrm{background}}).
\end{eqnarray}  
We  search for the two emerging top quarks in semi-leptonic (or semi-hadronic) 
modes, i.e., 
\begin{equation}
\label{decay}
t\bar t\to b\bar bW^+W^-\to b\bar b {\mathrm{j}}{\mathrm{j}}
\ell\nu,
\end{equation}
where j identifies a light-quark jets and $\ell=e,\mu$. We assume 
four $b$-tags but no $b$-jet
 charge determination, so that the final signature is:
\begin{equation}\label{signature}
4b+{\mathrm{j}}{\mathrm{j}}+\ell+p_T^{\rm{miss}},
\end{equation}
as the neutrino eventually escapes detection.

However, one can actually reconstruct the longitudinal momentum of the
neutrino, even in presence of Initial State Radiation (ISR). The method
is rather simple and it was initially outlined for the case of 
hadron-hadron collisions \cite{hadron}, where the initial boost due to 
the (anti)quarks
and gluons scattered out of the (anti)protons can have far more 
severe effects on the momentum reconstruction than those due to ISR in 
electron-positron annihilations. We will outline the procedure below.

In our numerical results, we
have assumed the MSSM throughout with 
$M_{A}$ ranging between 120 and 660 GeV.
Most of the signal plots will be presented for $\tan\beta=40$. However,
we will discuss other $\tan\beta$  values at the very end. For the signal, 
we have used the same program employed in \cite{singleHpm} 
for the production process and the one described in
\cite{BRs} for the Branching Ratios (BRs), 
all allowing for the inclusion of off-shellness
effects of the charged Higgs bosons\footnote{For the computation of
the backgrounds we have used
MadGraph \cite{MadGraph}.}.   The other unstable particles
entering the two processes, i.e., $t$ and $W^\pm$, were also generated
off-shell, with 
$\Gamma_t=1.55$ GeV and
$\Gamma_W=2.08$ GeV, in correspondence of
$m_t=175$ GeV  and
$M_{W^\pm}=M_Z\cos\theta_W\approx80$ GeV 
($M_Z=91.19$ GeV and $\sin^2\theta_W=0.232$). 
The non-running $b$-quark mass adopted for both the kinematics and the 
Yukawa coupling was $m_b=4.25$ GeV. We neglect ISR
and beamstrahlung effects, as we expect these to have a marginal
impact on the {\sl relative} behaviour of signal and background.

The entire simulation has been carried out at parton level, by identifying
jets with the partons from which they originate, though 
 finite calorimeter resolution
has been emulated through a Gaussian smearing in transverse momentum,
$p_T$, with $(\sigma(p_T)/p_T)^2=(0.60/\sqrt{p_T})^2 +(0.04)^2$
for jets and 
            $(\sigma(p_T)/p_T)^2=(0.12/\sqrt{p_T})^2 +(0.01)^2$
for leptons. The resulting missing transverse momentum, 
$p_T^{\mathrm{miss}}$, was reconstructed from the vector sum of 
the visible jet/lepton  momenta after resolution smearing. Finally, 
the integration over the final states 
has been performed numerically with the aid of VEGAS \cite{VEGAS}
and Metropolis \cite{Metro}.

After selecting the missing neutrino, the lepton, the 
$b$- and light-quark jets in the detector region, by imposing the 
following (acceptance) cuts in transverse momentum, polar angle
and cone separation:
$$
p_T^{\rm{miss}},~p_T^\ell,~p_T^b,~p_T^{~\rm{j}}>5~{\rm{GeV}},
$$
$$
|\cos\theta_\ell|,~
|\cos\theta_{b}|,~
|\cos\theta_{\rm{j}}|<0.995,
$$
\begin{equation}\label{acceptance}
\Delta R_{\ell b},~
\Delta R_{\ell{\rm{j}}},~
\Delta R_{bb},~
\Delta R_{b{\rm{j}}},~
\Delta R_{{\rm{j}}{\rm{j}}}~>0.4,
\end{equation}
one proceeds as follows\footnote{The adoption of a jet-clustering algorithm
\cite{schemes} instead of a cone one, as done in \cite{taunu,Battaglia},
would not affect our final conclusions.}.
\begin{itemize}
\item The invariant mass of the two non-$b$-jets is required
to be consistent with $M_{W^\pm}$,
\begin{equation}\label{MW}
|M_{\rm{jj}} - M_{W^\pm} | \le  15 \ {\rm GeV}.
\end{equation}
\item The invariant mass formed by combining the untagged jet
pair with one of the four $b$-jets is required to match $m_t$,
\begin{equation}
| M_{b\rm{jj}} - m_t | \le 25 \ {\rm GeV}.
\end{equation}
If several $b$-jets satisfy this constraint, the one giving the best
agreement with $m_t$ is selected.
\item The neutrino momentum is reconstructed by equating
$p_T^\nu = p_T^{\rm miss}$ and deducing the longitudinal component
$p_L^\nu$ from the invariant mass constraint $M_{\ell\nu} = M_{W^\pm}$.  The
resulting equation is quadratic, hence it can give
give two solutions.  If they are complex we discard their
imaginary parts so that they coalesce. Otherwise, both solutions are
retained. 
\item The invariant mass formed by combining $\ell$ and $\nu$
with one of the three remaining $b$-jets is also required to 
reproduce $m_t$:
\begin{equation}\label{mt}
| M_{b\ell\nu} - m_t | \le 25 \ {\rm GeV}.
\end{equation}
Again, if several $b$-jets satisfy the above requirement, the
one giving the best agreement with $m_t$ is selected along with the
corresponding value of $p_L^\nu$.
\item The remaining pair of $b$-jets may be looked upon as the
$b\bar b$ pair accompanying the $t\bar t$ in the signal (\ref{signal}) and
background (\ref{background}).  
Note that one of these $b$-jets is expected to come
from the $H^\pm$ decay in the signal, while for the background they both
come from a gluon splitting.  Consequently, in the latter case one supposes 
the $b\bar b$ pair to have a rather different kinematics
with respect to the former. We will eventually verify and make use of such
differences in order to optimise our selection\footnote{Note the more
conservative constraints adopted here in the $W^\pm$- and $t$-mass 
reconstruction
with respect to Ref.~\cite{taunu}, which is 
justified by the larger hadronic
multiplicity of the present final state.}.
\end{itemize}
 
\begin{figure}
\begin{center}
\epsfig{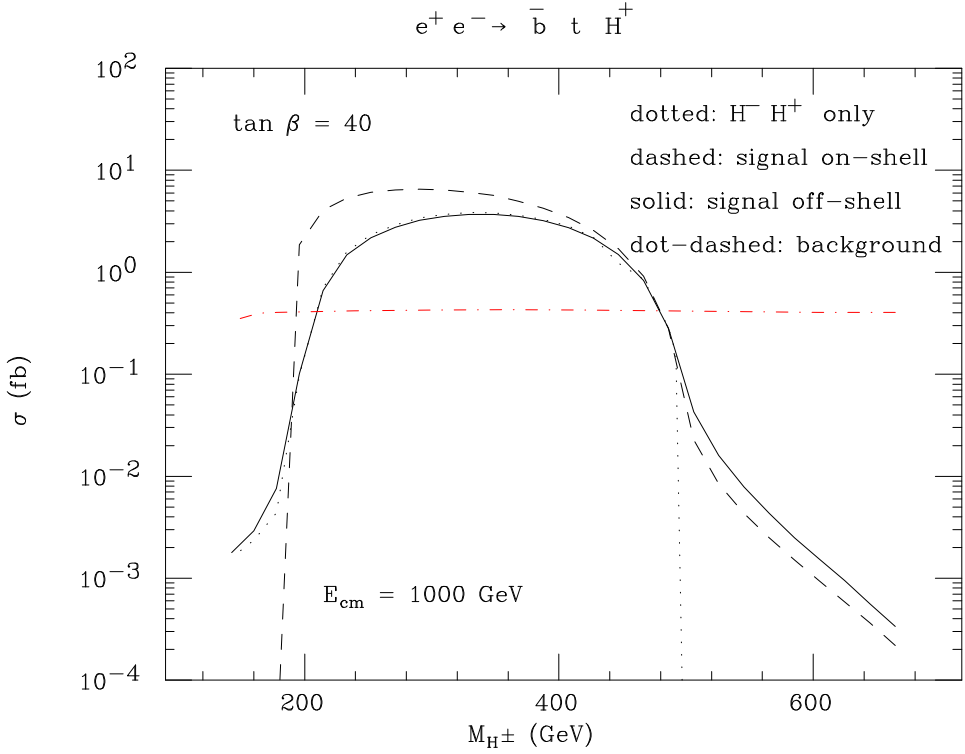}\\
\end{center}
\vspace{-0.5cm}
\caption{Total cross sections for processes (\ref{signal}) and  (\ref{background})
 yielding the signature (\ref{signature}), with 
the charged Higgs boson (and other unstable particles)
being on- and off-shell, including 
all decay BRs. No cuts have been 
enforced here. We also show the cross sections corresponding to graphs
proceeding via $e^+e^-\to H^-H^+$.}
\label{fig:off-shell}
\end{figure}

We start our numerical investigation by comparing the LC rates for 
process (\ref{proc_t}) computed when all unstable particles ($t$, $W^\pm$
and $H^\pm$) are set on-shell, i.e.,
 in Narrow Width Approximation (NWA), to those in 
which the latter are all allowed to be off-shell. 
The corresponding curves are displayed in
Fig.~\ref{fig:off-shell}. 
For reference, in the same figure, we also show the rates
obtained by using the two-body mode $e^+e^-\to H^-H^+$ followed by
the decay $H^-\to b \bar t$  (including off-shell
top (anti)quarks). No cuts are enforced here.
At the `threshold' point $M_{H^\pm}\approx
\sqrt s/2$, one may notice that two of the 
curves start departing. These contain
diagrams other than those proceeding via
$e^+e^-\to H^-H^+$ as well as the relative interference
between the two sets of graphs. The difference
between the two curves is due to the
finite width of the charged Higgs boson, which is of 10 GeV or so at
500 GeV and above. At $M_{H^\pm}\approx m_t$, one may also appreciate
the effects of finite values of 
$\Gamma_t, \Gamma_{W^\pm}$ and $\Gamma_{H^\pm}$.
We are however interested in the $M_{H^\pm}\gsim\sqrt s/2$ region.
Here, one should notice that
the $t\bar tb\bar b$ background   (including the decay
BRs yielding the final state in (\ref{signature})) is about 0.40 fb
(at leading order)\footnote{Here, the small drop for $M_{H^\pm}$
below $m_t$ is due to the fact that we have not included
$t\to bH^+$ decays in the definition of the background.}. 
The $S/B$ ratio is rather small then, to
start with, about 1/20 at $M_{H^\pm}\approx \sqrt s/2$ (recall that
$\sqrt s=1000$ GeV), yet not prohibitive to attempt disentangling
the signal from the irreducible background.

\begin{figure}
\begin{center}
\epsfig{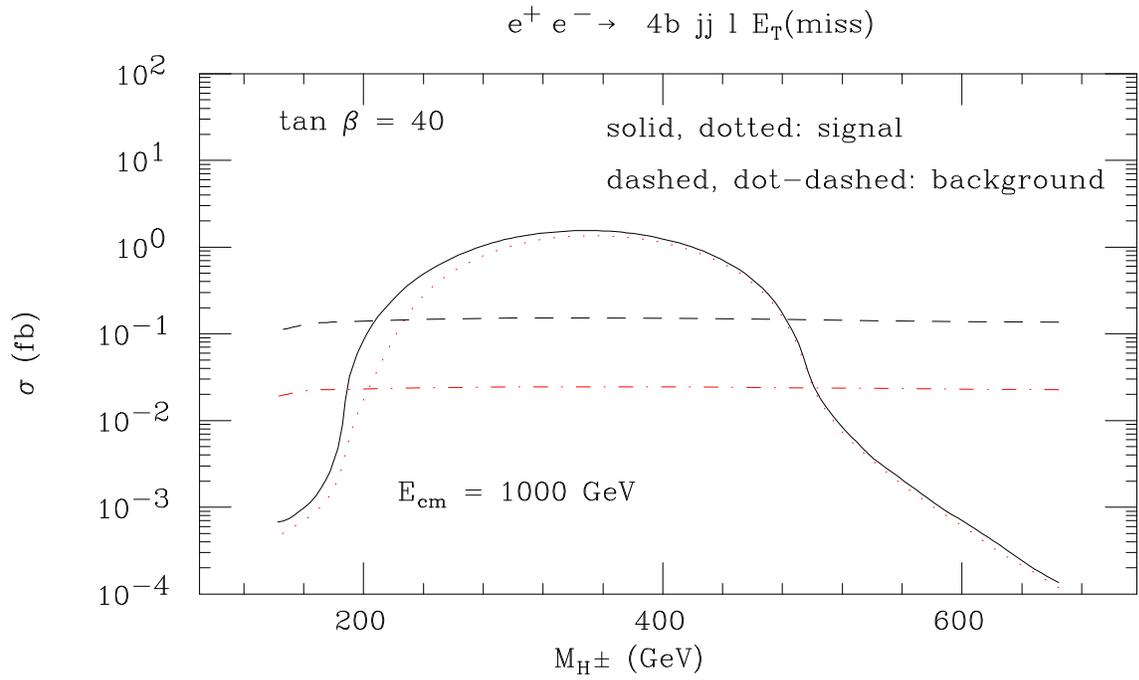}\\
\end{center}
\vspace{-0.5cm}
\caption{Total cross sections for processes 
(\ref{signal}) and (\ref{background}) yielding the signature 
(\ref{signature}),
after the kinematic cuts in (\ref{acceptance})--(\ref{mt})
[solid and dashed lines] and after the additional 
cuts in (\ref{bbcuts}) [dotted and dot-dashed lines], including 
all decay BRs.}
\label{fig:cuts}
\end{figure}

We now proceed in our investigation by enforcing the acceptance and
selection cuts in eqs.~(\ref{acceptance})--(\ref{mt}). The signal
and background cross sections which survive these constraints can
be found in Fig.~\ref{fig:cuts} (solid and dashed lines). Despite also the 
background contains two top (anti)quarks, the improvement in  $S/B$ is 
substantial, as, at $M_{H^\pm}\approx\sqrt s/2$, it has now decreased to
1/8 or so. The main cause for the severe 
background reduction turns out to be the
separation cuts between jets, as in the $g^*\to b\bar b$ splitting the
two emerging $b$-jets tend to be collinear\footnote{Note that the
gluon radiated off the $t\bar t$ pair can be rather energetic, as
$\sqrt s\gg 2m_t$.}.

\begin{figure}
\begin{center}
\epsfig{file=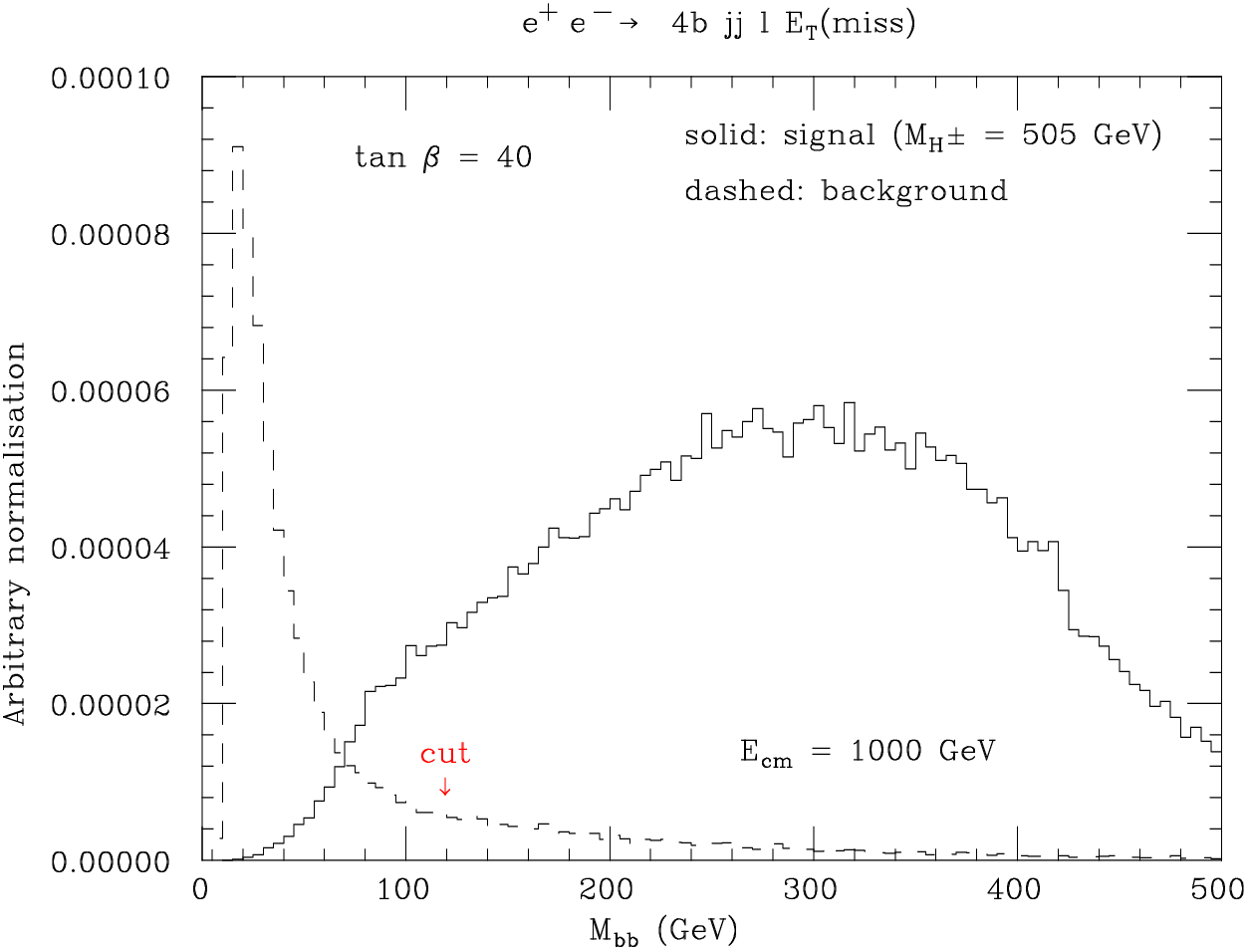,width=10cm,angle=90}\\
\end{center}
\vspace{-0.5cm}
\caption{Differential distribution in the invariant mass of the 
two $b$-jets not generated in top decays
for processes (\ref{signal}) and (\ref{background}) 
yielding the signature (\ref{signature}),
after the
kinematic cuts in (\ref{acceptance})--(\ref{mt}), including 
all decay BRs.}
\label{fig:Mbb}
\end{figure}

\begin{figure}
\begin{center}
\epsfig{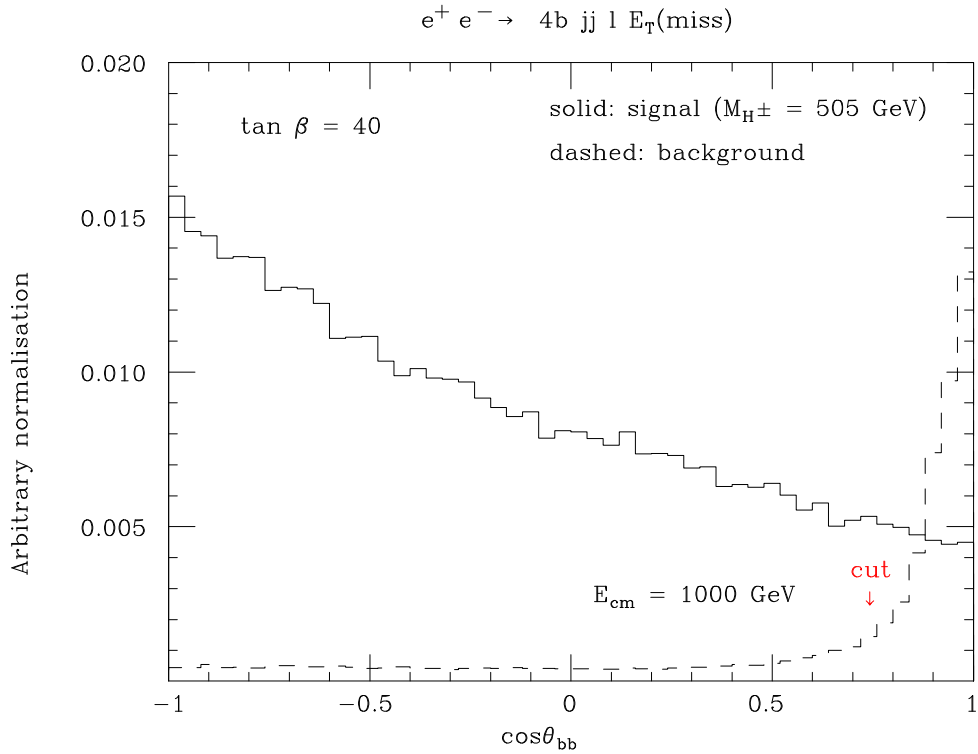}\\
\end{center}
\vspace{-0.5cm}
\caption{Differential distribution in the relative angle of the 
two $b$-jets not generated in top decays
for processes (\ref{signal}) and (\ref{background}) 
yielding the signature (\ref{signature}),
after the
kinematic cuts in (\ref{acceptance})--(\ref{mt}), including 
all decay BRs.}
\label{fig:cbb}
\end{figure}

\begin{figure}
\begin{center}
\epsfig{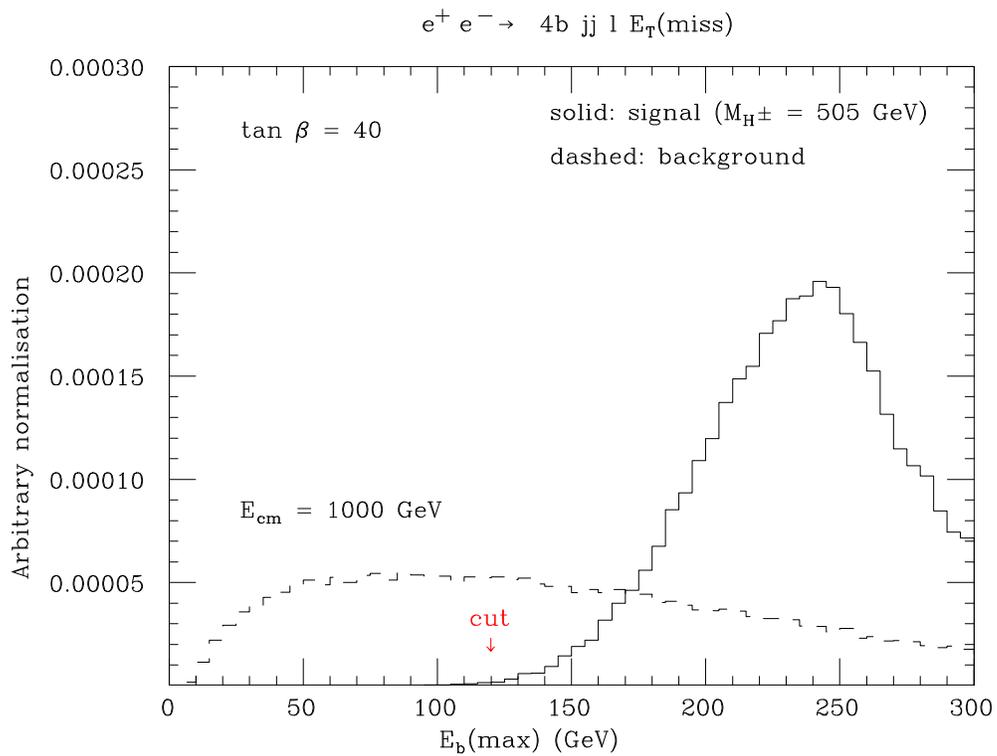}\\
\end{center}
\vspace{-0.5cm}
\caption{Differential distribution in energy of the hardest of the
two $b$-jets not generated in top decays
for processes (\ref{signal}) and (\ref{background}) 
yielding the signature (\ref{signature}),
after the
kinematic cuts in (\ref{acceptance})--(\ref{mt}), including 
all decay BRs.}
\label{fig:Ebmax}
\end{figure}

An even more vigorous reduction of the background rate, especially
in the very heavy Higgs mass region, can
be obtained by exploiting the different kinematic behaviour
of the hadronic system involving the two $b$-jets not associated
to the reconstructed pair of top (anti)quarks, as mentioned earlier on. 
Figs.~\ref{fig:Mbb}--\ref{fig:Ebmax} 
illustrates the dependence of signal and background rates
on the three following variables:
the invariant mass, $M_{bb}$, the (cosine of the) 
relative angle, $\cos\theta_{bb}$, and the energy of the hardest
of the two $b$-jets,  $E_{b}({\rm {max}})$. The Higgs mass in the
signal has been chosen in the critical region, $M_{H^\pm}\gsim\sqrt s/2$:
we have taken $M_{H^\pm}=505$ GeV as an illustration. Upon investigation 
of Figs.~\ref{fig:Mbb}--\ref{fig:Ebmax}, an effective combination of
cuts seems to be the following:
\begin{equation}\label{bbcuts}
M_{bb}>120~{\mathrm{GeV}},
\qquad
\qquad
\cos\theta_{bb}<0.75,
\qquad
\qquad
E_{b}({\rm{max}})>120~{\mathrm{GeV}}.
\end{equation}
(Note that such quantities are all correlated, so that the consequent effects 
do not factorise.)

The dotted and dot-dashed curves
in Fig.~\ref{fig:cuts} present the signal rates after the 
improved kinematic selection has been enforced, that is,
after the implementation of the constraints in 
eqs.~(\ref{acceptance})--(\ref{bbcuts}). (For $M_{H^\pm}>m_t$, the background 
cross section continues to be constant with $M_{H^\pm}$ as none of the
cuts used so far depends on this parameter.) The improvement in $S/B$ is
dramatical, as the signal remains basically unaffected by the additional
cuts, while the background is reduced by a factor of almost six.

\begin{figure}
\begin{center}
\epsfig{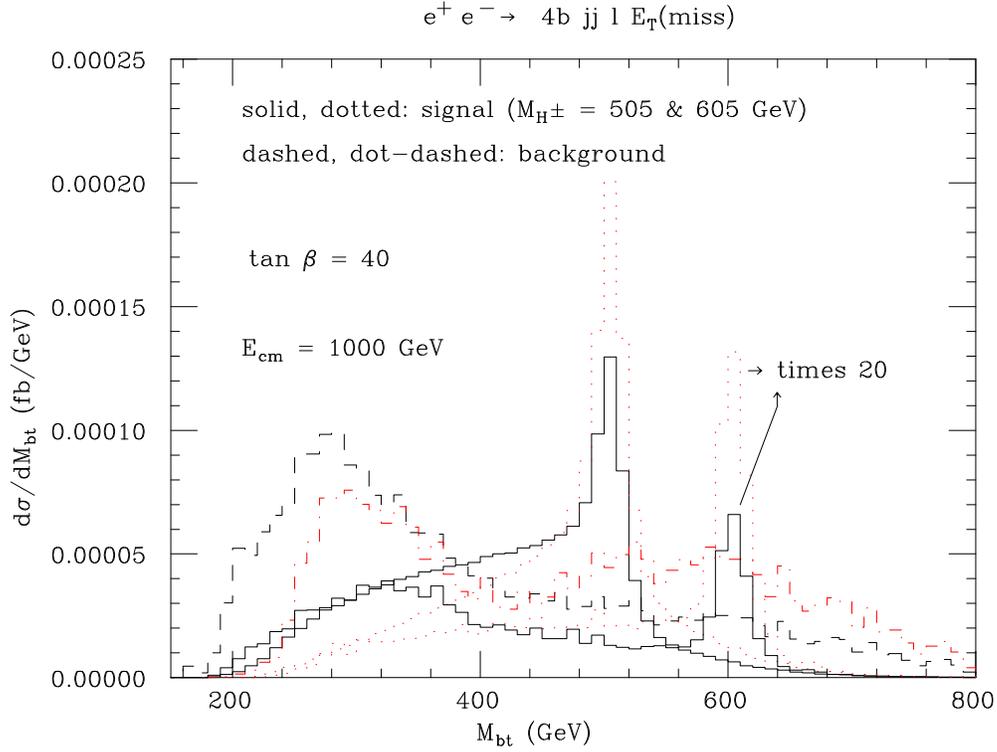}\\
\end{center}
\vspace{-0.5cm}
\caption{(Solid and dashed lines) 
Differential distribution in the reconstructed Higgs mass 
from both $b$-jets not generated in top decays and the two 
top systems (see definition (a) in the text)
for the processes (\ref{signal}) and (\ref{background})
yielding the signature (\ref{signature}),
after the
kinematic cuts in (\ref{acceptance})--(\ref{bbcuts}), including 
all decay BRs. (Dotted and dot-dashed lines) Same as above, but using only 
the $b$-jet with highest transverse momentum of the two 
not generated in top decays (see definition (b) in the text).}
\label{fig:MH}
\end{figure}

After the outlined procedure has been enforced, one is in the position
of being able to reconstruct the charged Higgs boson mass. There are
two possible ways to proceed in doing so.
\begin{enumerate}
\item[(a)] One can combine each of the reconstructed $t$-quarks
with each of the remaining $b$-jets to obtain four entries for the
$bt$ invariant mass, $M_{bt}$.  For each signal point, one of these
entries will correspond to the parent $H^\pm$ mass while the other
three will represent the combinatorial background.  We  plot the
signal and background cross sections against this quantity
in Fig.~\ref{fig:MH} (solid and dashed lines).  The
former clearly shows the resonant peak at $M_{H^\pm}$ emerging over
a broad combinatorial background, while the latter presents only a rather
flat distribution in $M_{bt}$ near the signal resonances.  
\end{enumerate}
For $M_{H^\pm} \gg m_t$, indeed the region of our interest,
one of the two spare $b$-jets in the signal
(namely, the one coming from the $H^+ \rightarrow t\bar b$ decay)
would generally be much harder than the other. Hence one 
can expect to reduce the combinatorial background as follows.
\begin{enumerate}
\item[(b)] One can combine each of the reconstructed $t$-quarks with
the harder of the two accompanying 
$b$-jets.  This gives two values for the invariant mass
$M_{bt}$ for each signal point, one of which corresponds to the
parent $H^\pm$ mass.  We show the signal and
background cross sections against this quantity 
in  Fig.~\ref{fig:MH} (dotted and dot-dashed lines). 
The pattern is similar to what
seen previously, yet the signal resonances are now
significantly narrower.
\end{enumerate}

In either case, despite the
width of the signal spectra is dominated by
detector smearing effects, it is clear that
the Breit-Wigner peaks themselves can help to improve 
$S/B$ further as well as to determine the $H^\pm$ mass. 
Fig.~\ref{fig:MH} suggests then that a further selection
criterium can be afforded at this stage: e.g.,
\begin{equation}\label{btcut}
|M_{bt}-M_{H^\pm}|<40~{\rm GeV}. 
\end{equation}
The value of $M_{H^\pm}$ entering eq.~(\ref{btcut}) would be the central 
or fitted mass resonance of the region in $M_{bt}$ where an excess of the form 
seen in Fig.~\ref{fig:MH} will be established. We perform the exercise
for both mass spectra (the one involving four entries and the one 
using two).

\begin{figure}
\begin{center}
\epsfig{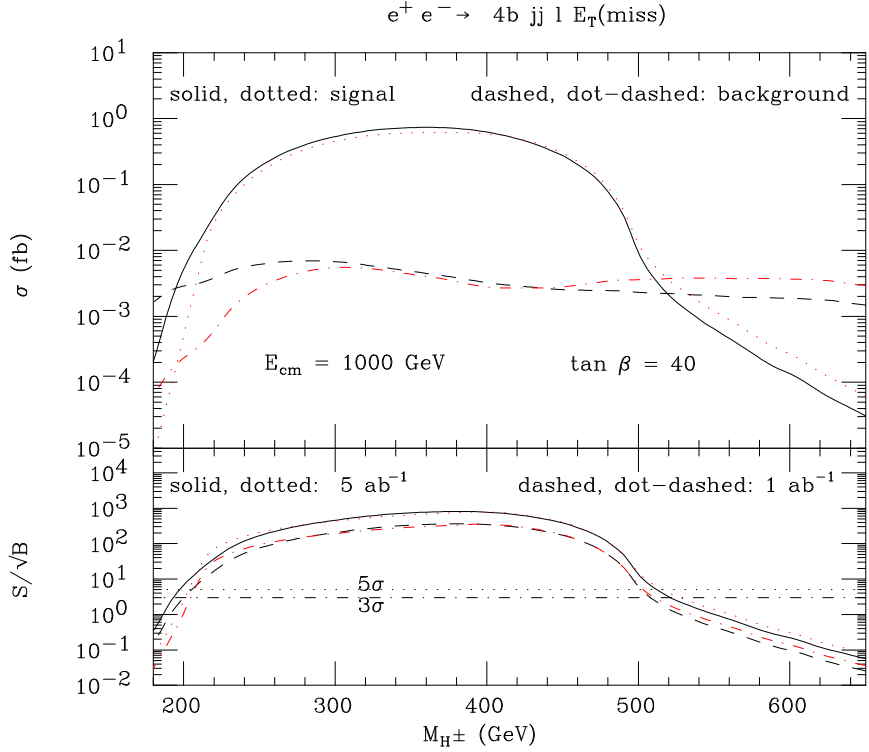}\\
\end{center}
\vspace{-0.5cm}
\caption{(Top) Total cross sections for processes 
(\ref{signal}) and (\ref{background}) yielding the signature 
(\ref{signature}),
after the kinematic cuts in (\ref{acceptance})--(\ref{bbcuts}) plus the one in
(\ref{btcut}) applied to the $M_{bt}$ distribution defined in (a) 
[solid and dashed lines] and in (b) [dotted and dot-dashed lines] 
(see the text), 
including all decay BRs. (Bottom) Corresponding 
statistical significances of the signal for two values of integrated
luminosity (the $3\sigma$ and $5\sigma$ `evidence' and `discovery' 
thresholds are also given).
}
\label{fig:stat}
\end{figure}

The resulting cross sections for signal and background, after the
additional cut in (\ref{btcut}), are shown in the top frame of 
Fig.~\ref{fig:stat}, as obtained from the two $M_{bt}$ distributions above.
The signal rates are not very large in the $M_{H^\pm}\gsim \sqrt s/2$
region with the background ones being comparable or even
higher. Yet, with the very large luminosity that can be accumulated at a 
future LC, statistical significances ($S/\sqrt B$) may be sizable.
The latter are displayed, after 1 and 5 ab$^{-1}$ of accumulated
luminosity, $\cal L$, in the bottom frame of 
Fig.~\ref{fig:stat}, again,  as obtained from the two 
definitions of $M_{bt}$ given in (a) and (b).
(The first luminosity figure may well be attained according to
current designs, while the second should be viewed  at present as an 
optimistic scenario.) By comparing these curves with the shape of the
$e^+e^-\to H^-H^+$ cross
section in Fig.~\ref{fig:off-shell}, one should expect to extend the reach 
in $M_{H^\pm}$ obtained from pair production of charged Higgs bosons and 
decays by at most 25 GeV at $3\sigma$ level above the $M_{H^\pm}= \sqrt s/2$
point and only at very high luminosity, thanks to the contribution of
single $H^\pm$ production in association with top- and bottom-(anti)quarks. 
Typical signal rates at $\tan\beta=40$ in the threshold
region $M_{H^\pm}\approx\sqrt s/2$
would be about 5(25) events for definition (a) of $M_{bt}$
and 7(35) for (b), in correspondence of $\cal L=$ 1(5) ab$^{-1}$.

\begin{figure}
\begin{center}
\epsfig{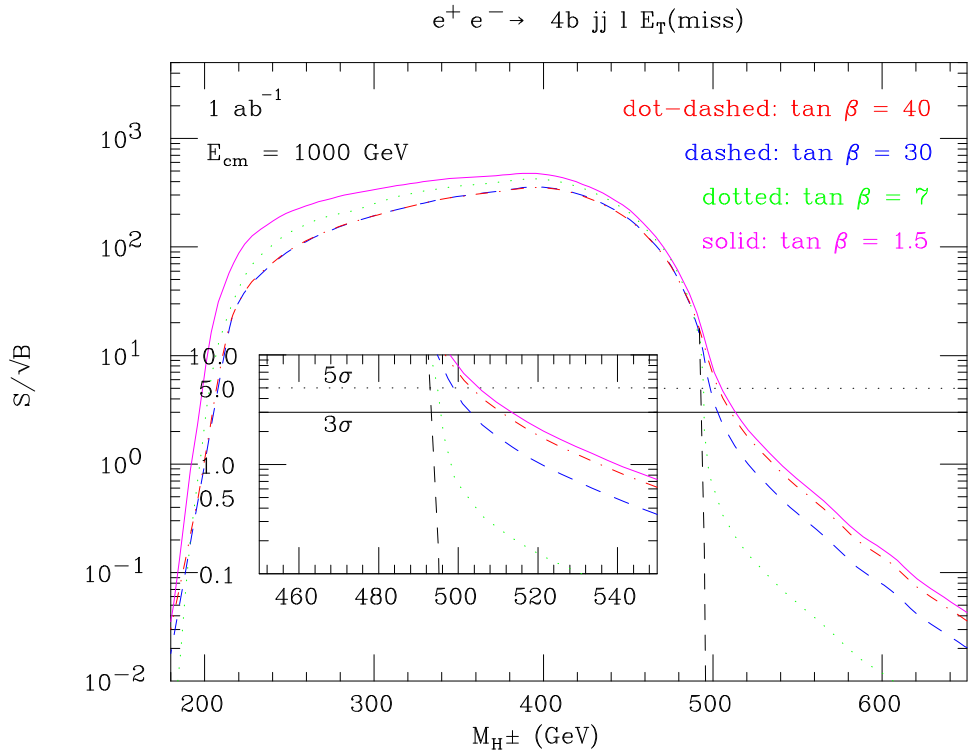}\\
\end{center}
\vspace{-0.5cm}
\caption{Statistical significances of the signal (\ref{signal}) yielding the signature (\ref{signature}),
after the kinematic cuts in (\ref{acceptance})--(\ref{bbcuts}) plus the one in
(\ref{btcut}) applied to the $M_{bt}$ distribution defined in (b) (see
the text), including 
all decay BRs, for four different values of $\tan\beta$.
The value of integrated
luminosity is here 1 ab$^{-1}$
(the $3\sigma$ and $5\sigma$ `evidence' and `discovery' 
thresholds are also given).
The vertical dashed line corresponds to the kinematic limit of
the graphs proceeding via $e^+e^-\to H^-H^+$ (see also Fig.~1).
In the insert, we enlarge the region $M_{H^\pm}\sim  \sqrt s/2$.
}
\label{fig:tanB}
\end{figure}

However, recall that process (\ref{proc_t}) is very sensitive to the actual 
value of $\tan\beta$. While in the $\tan\beta\gsim40$ region
its production rates approximately
scale like $\tan\beta^2$ (so that the higher this parameter the better
the chances of isolating the signal discussed here, though 
a natural upper limit for it is expected at around $50$), for  $\tan\beta$
much smaller than 40 the statistical significance 
would diminish very rapidly, unless $\tan\beta$ values of
order 1.5 or so are allowed within the underlying model.
In fact, the average strength of the $t\bar b H^-$ coupling is
proportional to $\sqrt{m_t^2\cot\beta^2+m_b^2\tan\beta^2}$, which
has a minimum at $\tan\beta\approx\sqrt{m_t/m_b}\approx 6.4$.
Fig.~\ref{fig:tanB} illustrates this trend, for the more realistic
value of luminosity, ${\cal L}=1$ ab$^{-1}$, and limitedly to the 
kinematically
more favourable case in which $M_{bt}$ is computed as described in case (b) 
above. Under these conditions, the maximum mass reach at $3\sigma$ 
level beyond the
threshold region $M_{H^\pm}\sim \sqrt s/2$ is of 20 GeV or so, for
very large or very small $\tan\beta$. Finally, in our estimates so far, 
we have excluded the efficiency of tagging the four $b$-jets
in the final state. According to Ref.~\cite{Battaglia}, the single
$b$-tag efficiency is expected to be close to the value $\epsilon_b=90\%$, 
thus yielding -- if all four $b$-quarks are tagged --
a reduction factor of $\epsilon_b^4\approx0.66$ for the
number of events of both signal and background and of 
$\epsilon_b^2\approx0.81$ for the statistical significance $S/\sqrt B$,
hence diminishing even further the scope of our signature. Eventually, 
the impact of the next-to-leading order 
QCD corrections computed in Ref.~\cite{Bernd}
should also be evaluated carefully, as they can alter the
leading order results for the signal by up to $\pm50\%$,
depending on the values of $\sqrt s$, $M_{H^\pm}$ and $\tan\beta$. We refrain
from doing so here for consistency, as the irreducible background that
we have considered would also be subject to large QCD effects, which
are however still unknown. 

In summary, while not been very encouraging per se,  
if combined with the results of Refs.~\cite{singleHpm}--\cite{WH} and
\cite{taunu,Shingo} 
for the $\tau^-\bar\nu_\tau H^+$ and $W^-H^+$ 
channels, our present findings
should be taken into account in establishing strategies to detect 
heavy charged Higgs bosons with masses $M_{H^\pm}\gsim \sqrt s/2$
at future LCs, especially considering
the very high level of background reduction achieved through
our kinematic selection, which started from very poor 
$S/B$ rates. Unfortunately, absolute rates
for singly produced heavy $H^\pm$ states are never very large either,
so that to achieve high luminosity is of paramount importance.
Our conclusions have been derived within the MSSM, nonetheless,
they are equally applicable to a general Type II 2HDM, as the only
couplings involved in the present analysis are common to both scenarios.
In particular, in a 2HDM the mentioned
low $\tan\beta$ region is not excluded, so that also in this
framework the  $b\bar t H^+$ channel may well turn out to be of
use.

Two foreseen outlooks are the following. On the one hand, one can
attempt to exploit process (\ref{proc_wh}) in order to cover the
same kinematic region at low to intermediate 
$\tan\beta$ \cite{newpreparation}. 
On the other hand, one 
ought to eventually fold the parton level
 results into a more sophisticated simulation, 
as we are planning to do \cite{preparation}
in the context 
of the HERWIG Monte Carlo event generator \cite{HERWIG,me},
where more realistic estimates for $b$-tagging efficiency and
rejection against mis-identification of light-quark and gluon
jets would be obtained. In this context, one may also consider
the possibility of tagging three $b$-jets only, which would
still allow for the implementation of the kinematic selection
outlined here, at the same time increasing somewhat both the
signal rates and significances while not implying additional
backgrounds,
as $b$-quarks are always produced in pairs in $e^+e^-$ annihilations.

\vskip0.5cm
\noindent
\underbar{\sl Acknowledgements:}~We thank Bernd Kniehl for encouragement
to pursue this work and for illuminating discussions.

\clearpage

\clearpage

\clearpage

\clearpage

\clearpage

\clearpage

\clearpage

\clearpage

\end{document}